\documentclass[superscriptaddress,aps,prc]{revtex4}
\usepackage{graphicx}
\usepackage{dcolumn}
\begin{document}

\title{New $^{32}$Cl(p,$\gamma$)$^{33}$Ar reaction rate for astrophysical
rp-process calculations}

\author{H. Schatz}
\affiliation{National Superconducting Cyclotron Laboratory, Michigan State University,
East Lansing, MI 48824, USA}
\affiliation{Dept. of Physics and Astronomy, Michigan State University, East Lansing,  MI 48824, USA}
\affiliation{Joint Institute for Nuclear Astrophysics, Michigan State University, East Lansing, MI 48824, USA}

\author{C. A. Bertulani \footnote{Current Affiliation: Dept. of Physics, University of Arizona, Tucson, AZ 85721, USA}}
\affiliation{National Superconducting Cyclotron Laboratory, Michigan State University,
East Lansing, MI 48824, USA}

\author{B. A. Brown}
\affiliation{National Superconducting Cyclotron Laboratory, Michigan State University,
East Lansing, MI 48824, USA}
\affiliation{Dept. of Physics and Astronomy, Michigan State University, East Lansing,  MI 48824, USA}

\author{R. R. C. Clement \footnote{Current Affiliation: Lawrence Livermore National Laboratory, 7000 East Ave. Livermore, CA 94550, USA}}
\affiliation{National Superconducting Cyclotron Laboratory, Michigan State University,
East Lansing, MI 48824, USA}
\affiliation{Dept. of Physics and Astronomy, Michigan State University, East Lansing,  MI 48824, USA}

\author{A. A. Sakharuk}
\affiliation{National Superconducting Cyclotron Laboratory, Michigan State University,
East Lansing, MI 48824, USA}
\affiliation{Joint Institute for Nuclear Astrophysics, Michigan State University, East Lansing, MI 48824, USA}

\author{B. M. Sherrill}
\affiliation{National Superconducting Cyclotron Laboratory, Michigan State University,
East Lansing, MI 48824, USA}
\affiliation{Dept. of Physics and Astronomy, Michigan State University, East Lansing,  MI 48824, USA}
\affiliation{Joint Institute for Nuclear Astrophysics, Michigan State University, East Lansing, MI 48824, USA}

\begin{abstract}
The $^{32}$Cl(p,$\gamma$)$^{33}$Ar reaction rate is of potential importance in the rp-process powering type I 
X-ray bursts. Recently Clement et al. \cite{CBB04} presented new data on excitation energies for
low lying proton unbound states in $^{33}$Ar obtained with a new method developed at the National 
Superconducting Cyclotron Laboratory. We use their data, together with
a direct capture model and a USD shell model calculation to derive a new reaction rate for use in 
astrophysical model calculations. In particular, 
we take into account capture on the first excited state in $^{32}$Cl, and also present a realistic 
estimate of the remaining uncertainties. We find that the $^{32}$Cl(p,$\gamma$)$^{33}$Ar reaction 
rate is dominated entirely by capture on the first excited state in $^{32}$Cl over the whole
temperature range relevant in X-ray bursts. In the temperature range from 0.2 to 1 GK
the rate is up to a factor of 70 larger than 
the previously recommended rate based on shell model calculations only. The 
uncertainty is now reduced from up to a factor of 1000 to 
a factor of 3 at 0.3-0.7 GK and a factor of 6 at 1.5~GK. 
\end{abstract}



\maketitle

\section{Introduction}

Proton capture rates in the rapid proton capture process (rp-process) play a 
critical role in determining energy release and final isotopic 
abundances in X-ray bursts \cite{WaW81,SAG97,WiS99,SBB99,TBF01,FTW04}. 
Reliable rates
are therefore important for quantitative interpretations of 
observations. For example, new highly accurate data on burst 
profile changes over periods of years as observed in GS 1826-24 \cite{GCK04}
would provide unique constraints on X-ray burst 
models if the nuclear physics of the rp-process
would be well enough understood. As demonstrated
in a number of X-ray burst model calculations
\cite{KHA99,BCS01,WHC04} this is clearly not the case. 

One problem is the reliability of proton capture rates in the 
rp-process, especially 
in X-ray bursts where the reaction path runs close to the proton 
drip line. Far from stability, proton capture rates typically are governed by 
a few resonances and therefore statistical models are not 
applicable over the entire relevant temperature range of 
0.2-2~GK \cite{RTK97}.
On the other hand,
shell model calculations in the sd- \cite{BaW88,HGW95} and fp-shells \cite{HOB02,HOB04,FBG01}
predict the properties of individual states but the typical
uncertainty in level energies is about 100~keV even for the calculations of 
level shifts from mirror states. This
translates into uncertainties of 3 or more orders of magnitude
in the reaction rates \cite{SGH97,CBB04}. Currently, we expect that the majority of the 
proton capture rates in the rp-proces suffer from such uncertainties. 

Clement et al. \cite{CBB04} recently developed a new experimental method 
using radioactive beams
at the National Superconducting Cyclotron Laboratory at Michigan State
University to accurately
determine excitation energies of nuclei in the rp-process path. 
They presented new results for the excitation energies of states in 
$^{33}$Ar that determine the $^{32}$Cl(p,$\gamma$)$^{33}$Ar reaction 
rate. Based on the new experimental data we present here 
a reevaluation of the $^{32}$Cl(p,$\gamma$)$^{33}$Ar rate, which is
part of the reaction flow through the $^{30}$S - $^{34}$Ar region,
an important bottleneck 
in the rp-process in X-ray 
bursts possibly related to the observed double peaked structure 
of some X-ray burst profiles \cite{FTW04}. 

The $^{32}$Cl(p,$\gamma$)$^{33}$Ar reaction rate recommended
in reaction rate compilations \cite{Iliadis01} and used
in X-ray burst models so far was a ground state capture rate based on 
shell model calculations with some experimental information from the $^{33}$Ar mirror 
nucleus $^{33}$P \cite{HGW95}. 
We use the new experimental data on states in $^{33}$Ar together 
with shell model calculations and a direct capture model 
to derive a new dramatically improved rate for the
$^{32}$Cl(p,$\gamma$)$^{33}$Ar reaction. This differs from the 
initial discussion in Clement et al. \cite{CBB04} as we also take into 
account the contribution from proton capture on the first 
excited state in $^{32}$Cl at 89.9~keV. This state
is thermally populated in an astrophysical plasma and as we 
will show here actually dominates the proton capture rate
on $^{32}$Cl for the important temperature range between
0.2 and 1.5~GK. We also implement slight improvements in the 
penetrability calculations, evaluate the remaining shell model
uncertainties and present the reaction rate in a form usable 
by astrophysical reaction networks. 

\section{Resonant capture}

The resonant reaction rate for capture on a nucleus in an initial state $i$,
$N_A <\sigma v>_{ {\rm res}\, i}$ can for isolated narrow resonances be calculated 
as a sum over all relevant compound nucleus states $j$ above the proton threshold \cite{FoH64}:
\begin{equation} \label{EqRes}
N_A <\sigma v>_{{\rm res}\, i} = 1.540 \times 10^{11} (\mu T_9)^{-3/2} \sum_j 
\omega\gamma_{ij} {\rm e}^{-E_{ij}/(kT)} \, \, \, {\rm cm^3 \, s^{-1} mole^{-1}}
\end{equation}
where the resonance energy in the center of mass system $E_{ij} = E_j - Q - E_i$ 
is calculated from the excitation energies of the initial $E_i$ and 
compound nucleus $E_j$ state. $k$ is the Boltzmann constant. 
We adopt an experimental ground state mass difference $Q$ (reaction Q-value)
of 3.343$\pm$0.007 MeV \cite{AME03}.
$T_9$ is the temperature in GK and $\mu$ is the reduced mass
of the entrance channel in amu. The resonance strengths $\omega \gamma_{ij}$
are in MeV and can be calculated for proton capture as
\begin{equation} \label{EqOG}
\omega \gamma_{ij} = \frac{2J_j+1}{2(2J_i+1)} \frac{\Gamma_{{\rm p}\, ij} \Gamma_{\gamma j}}
{\Gamma_{{\rm total}\, j}}  
\end{equation}
where $J_i$ is the target spin, and $J_j$, $\Gamma_{{\rm p} \, ij}$, $\Gamma_{\gamma \,j}$,
${\Gamma_{{\rm total}\,j}}$
are spin, proton decay width, $\gamma$-decay width, and total width of the compound nucleus state $j$.
The total width is given by $\Gamma_{{\rm total}\,j}=\Gamma_{\gamma \,j}
+ \sum_i \Gamma_{{\rm p} \, ij}$. 
The proton decay widths depend exponentially on the resonance energy and can be calculated
from the proton spectroscopic factor $C^2S_{ij}$ and the single particle proton width 
$\Gamma_{{\rm sp} \, ij}$ as $\Gamma_{{\rm p} \, ij} = C^2S_{ij} \Gamma_{{\rm sp} \,ij}$.
Here we calculated spectroscopic factors with the USD shell model as in 
Herndl et al. \cite{HGW95}. The single particle proton widths
for most states were calculated
from an exact evaluation of the proton scattering cross section from a Woods-Saxon
potential well. This method is more precise than the penetrability model used by 
Herndl et al. \cite{HGW95} and Clement et al. \cite{CBB04}, but agrees with previous work within 20\%.
The $\gamma$ widths $\Gamma_\gamma$ were
taken from Herndl et al. \cite{HGW95}. This is justified as new experimental level energies are only 
available for the lower lying resonances where $\omega \gamma$ is largely independent of $\Gamma_\gamma$ as
$\Gamma_\gamma >> \Gamma_{\rm p}$, and because for those states the 
changes in the $\gamma$-widths are less than 20\%. 
The resulting properties of 
the resonance states in $^{33}$Ar are listed in Tab.~\ref{TabRes}.

\begin{table}
\caption{\label{TabRes}Properties of resonant states. Listed are spin and parity $J^\pi$, 
           excitation energy $E_x$, center of mass resonance energy $E_r$, proton single particle widths $\Gamma_{\rm sp}$ for angular 
           momenta $l$, spectroscopic factors $C^2S$, proton-decay width $\Gamma_p$, $\gamma$-decay width $\Gamma_\gamma$ and the 
           resonance strength $\omega\gamma$. The upper part is for ground state capture, the lower part for capture on 
           the first excited state in $^{32}$Cl.}
\begin{ruledtabular}
\begin{tabular}{cccccccccc}
$J^\pi$  & $E_x {\rm (MeV)}$ & $ E_r {\rm (MeV)}$ & \multicolumn{2}{c}{$\Gamma_{\rm sp}$}  & \multicolumn{2}{c}{$C^2S$} & $\Gamma_{\rm p} {\rm (eV)}$ & $\Gamma_\gamma {\rm (eV)}$ & $\omega\gamma {\rm (eV)}$ \\
 & & & l=0 & l=2 & l=0 & l=2 \\ 
$5/2^+$ & 3.364 & 0.021 $\pm$ 0.009 &           &  8.00e-42 &           &  2.90e-02 &  2.32e-43 &  1.77e-02 &  2.32e-43 \\ 
$7/2^+$ & 3.456 & 0.113 $\pm$ 0.009 &           &  1.90e-13 &           &  3.00e-03 &  5.70e-16 &  1.94e-03 &  7.60e-16 \\ 
$5/2^+$ & 3.819 & 0.476 $\pm$ 0.008 &           &  2.10e-02 &           &  4.30e-02 &  9.03e-04 &  1.54e-02 &  6.88e-04 \\ 
$1/2^+$ & 4.190 & 0.847 $\pm$ 0.100 &  6.30e+02 &  1.01e+01 &  6.70e-02 &  3.40e-02 &  4.26e+01 &  1.54e-01 &  5.11e-02 \\ 
$3/2^+$ & 4.730 & 1.387 $\pm$ 0.100 &  2.50e+04 &  6.30e+02 &  3.00e-03 &  3.90e-02 &  9.96e+01 &  8.48e-02 &  4.33e-03 \\ 
\hline\hline$7/2^+$ & 3.456 & 0.023 $\pm$ 0.009 &           &  1.30e-39 &           &  4.64e-03 &  6.03e-42 &  1.94e-03 &  4.83e-42 \\ 
$5/2^+$ & 3.819 & 0.386 $\pm$ 0.008 &  1.35e-01 &  1.50e-03 &  2.40e-02 &  4.39e-01 &  3.90e-03 &  1.54e-02 &  1.78e-03 \\ 
$1/2^+$ & 4.190 & 0.757 $\pm$ 0.100 &           &  3.50e+00 &           &  2.27e-03 &  7.94e-03 &  1.54e-01 &  5.73e-06 \\ 
$3/2^+$ & 4.730 & 1.297 $\pm$ 0.100 &  1.60e+04 &  3.80e+02 &  7.49e-02 &  3.77e-03 &  1.20e+03 &  8.48e-02 &  3.13e-02 \\ 
\end{tabular}
\end{ruledtabular}
\end{table}

\section{Direct capture}

The contribution to the $^{32}$Cl(p,$\gamma$)$^{33}$Ar rate from 
direct capture into bound states has been calculated with a
potential model \cite{bert03} using 
a Woods-Saxon nuclear potential (central+spin-orbit) and a Coulomb 
potential of a uniform charge distribution. 
The nuclear potential parameters were determined by 
matching the bound state energies. Spectroscopic factors
were calculated with the USD shell model as in 
Herndl et al. \cite{HGW95}. 

The direct proton capture 
capture rate $N_A<\sigma v>_{{\rm dc}\, i}$ on the target nucleus in state $i$
is usually parametrized in terms of the astrophysical
S-factor $S(E_0)$ in MeV~barn at the Gamov window energy $E_0$ \cite{FCZ67}:
\begin{equation}
N_A<\sigma v>_{\rm{dc}\, i} = 7.83 \times 10^9 \left( \frac{Z}{\mu T_9^2} \right) ^{1/3}
S_i(E_{o}) {\rm e}^{-4.29(Z^2 \mu/T_9)^{1/3}} \, \, \, {\rm cm^3 \, s^{-1} mole^{-1}}
\end{equation}
with $Z$ being the charge number of the target nucleus. $S_i(E_{0})$ is the 
sum of the individual S-factors of the transitions from the initial state $i$
into all bound states in the final nucleus. Table \ref{TabDC} lists the
individual S-factors found for transitions into the 5 bound states of 
$^{33}$Ar from both, the $^{32}$Cl ground state and the first excited state
as well as the total S-factor. For ground state capture, our total S-factor
agrees within 30\% with the one derived by Herndl et al. \cite{HGW95}
using the same spectroscopic factors. 

\begin{table}
\caption{\label{TabDC}Spectroscopic factors $C^2S$ and astrophysical S-factors $S(E_0)$ for direct capture into bound states in 
$^{33}$Ar. Listed are results for capture on the $^{32}$Cl ground state as well as on the first excited state in  $^{32}$Cl 
(denoted with an asterix). $J^\pi$ are spin and parity of the $^{33}$Ar final state, $n$ is the node number, $l_0$ the 
single particle orbital momentum, and $j_0$ the total single particle angular momentum. The table only includes
the significant contributions to the total S-factor. 
}
\begin{ruledtabular}
\begin{tabular}{dccdcdc}
 E_x {\rm (MeV)} & $J^\pi$ & $(nl_0)_{j0}$ & C^2S & $S(E_0) {\rm (MeV barn)}$ &  C^2S^* & $S(E_0)^* {\rm (MeV barn)}$ \\
0       & $1/2^+$ & 2s$_{1/2}$   & 0.08  & 6.56e-3                &           &         \\
        &         & 1d$_{3/2}$   & 0.67  & 4.47e-3                & 1.13      & 4.45e-3 \\
1.359   & $3/2^+$ & 2s$_{1/2}$   &       &                        & 0.006     & 6.61e-4 \\
        &         & 1d$_{3/2}$   & 0.19  & 1.78e-3                & 0.12      & 6.80e-4 \\ 
1.798   & $5/2^+$ & 2s$_{1/2}$   &       &                        & 0.002     & 3.21e-4 \\
        &         & 1d$_{3/2}$   & 0.15  & 1.84e-3                & 0.62      & 4.72e-3 \\
        &         & 1d$_{5/2}$   &       &                        & 0.021     & 1.65e-4 \\
2.439   & $3/2^+$ & 2s$_{1/2}$   & 0.031 & 4.32e-3                & 0.024     & 1.99e-3 \\
        &         & 1d$_{3/2}$   & 0.17  & 1.09e-3                & 0.13      & 5.01e-4 \\
3.154   & $3/2^+$ & 2s$_{1/2}$   & 0.068 & 1.02e-2                &           &         \\
        &         & 1d$_{3/2}$   & 0.52  & 2.21e-3                & 0.17      & 4.35e-4 \\ \hline
\mbox{total} &    &              &       & 3.25e-2                &           & 1.39e-2 \\
\end{tabular}
\end{ruledtabular}
\end{table}

\section{New reaction rate}

The relative population of the target nucleus states in thermodynamic equilibrium 
is simply given 
by the Saha equation. The total reaction rate is then the sum of the 
capture rate on all thermally excited
states in the target nucleus weighted with their individual 
population factors:

\begin{equation} \label{EqPop}
N_A <\sigma v> = \sum_i (N_A<\sigma v>_{{\rm res} \, i} +
                      N_A<\sigma v>_{{\rm dc}\, i} )
      \frac{(2J_i+1) {\rm e}^{-E_i/kT}}
                      {\sum_n (2J_n+1) {\rm e}^{-E_n/kT}}.
\end{equation}
To obtain the total resonant reaction rate on a thermally excited target one can 
combinine Eqs. \ref{EqRes}, \ref{EqOG}, and \ref{EqPop} and finds \cite{Van98}:
\begin{equation} \label{EqTotal}
N_A <\sigma v> = \sum_j N_A <\sigma v>_{0j} \frac{1}{G(T)} \left( 1 +
\sum_{i>0}^{E_{ij}>0} \frac{ \Gamma_{{\rm p} \, ij}}
{ \Gamma_{{\rm p} \, 0j}} 
\right)
\end{equation}
where $j$ sums over all resonances in the compound nucleus. $i$ sums over the 
thermally populated
states in the target nucleus as long as the resonance energy $E_{ij}>0$
with $i=0$ being the ground state. 
$N_A <\sigma v>_{0j}$ is the reaction rate contribution from ground state capture
via the resonance $j$. 
$G(T)$ is the
temperature $T$ dependent partition function of the target
nucleus:
\begin{equation}
 G(T)=(2J_0+1)^{-1} \sum_i (2J_i+1) {\rm exp}(-E_i/kT)
  \end{equation}
This is similar to Eq. 10 in 
Fowler, Caughlan and Zimmerman \cite{FCZ75}. 
Eq.~\ref{EqTotal} shows that for a given resonance the relative contribution from capture on each excited target state depends
only weakly on temperature and the actual population of the excited state through the partition function $G(T)$. The 
contribution of excited states is mostly determined by the ratio of the proton widths to the excited state and to 
the ground state. 
Of course the temperature determines through 
$<\sigma v>_{0j}$ the relative importance of the various resonances. 

In this work we consider only capture on the ground state and the first excited state in
$^{32}$Cl. The $^{32}$Cl ground state is experimentally known to have spin and
parity 1$^+$. The spin for the experimentally known first excited state
at 89.9~keV has not been determined unambiguously
but we assign a spin of $2^+$ based on the level structure of the $^{32}$P mirror and
our shell model calculations. Excited states in the target will only play a role when
for the relevant temperatures the thermal excitation timescale is smaller than the 
proton capture timescale. We can estimate the thermal excitation timescale for a
$\gamma$ induced transition from the ground state to the first excited $2^+$ state 
in $^{32}$Cl by using the formalism of Ward and Fowler \cite{WaF80}. We 
assume a level lifetime of 280~ps from the mirror
level in $^{32}$P, which is an upper limit as the mirror state is 12~keV lower
in excitation energy. 
We find that even for the lowest relevant temperatures of 0.2~GK
the excitation timescale is of the order of 30~ns. Assuming typical X-ray burst conditions
with a density of 10$^6$~g/cm$^3$ and a proton mass fraction of 0.7 the 
timescale for thermal excitation is always less than 0.2\% of the proton capture 
timescale for all temperatures between 0.2~GK and 2~GK. Therefore, the first 
excited 2$^+$ state in $^{32}$Cl is always in thermal equilibrium with the ground state
and Eq.~\ref{EqTotal} applies. 

What is the role of higher lying excited states in $^{32}$Cl? 
The second excited state in $^{32}$Cl is located at an energy of
466.1~keV. At such a high excitation energy
the proton decay width $\Gamma_{\rm p}$ in Eq.~\ref{EqTotal} is already drastically 
reduced for the low lying resonances that have a significant ground state
capture rate $<\sigma v>_{0j}$. Therefore, the capture rate on the second excited state 
in $^{32}$Cl is negliglible as either $\Gamma_{{\rm p} \, 2j}$ or 
$<\sigma v>_{0j}$ in Eq.~\ref{EqTotal} are small for all resonances. Similar
arguments apply to higher lying states. 

\begin{figure}[ht]
\includegraphics[width=8cm]{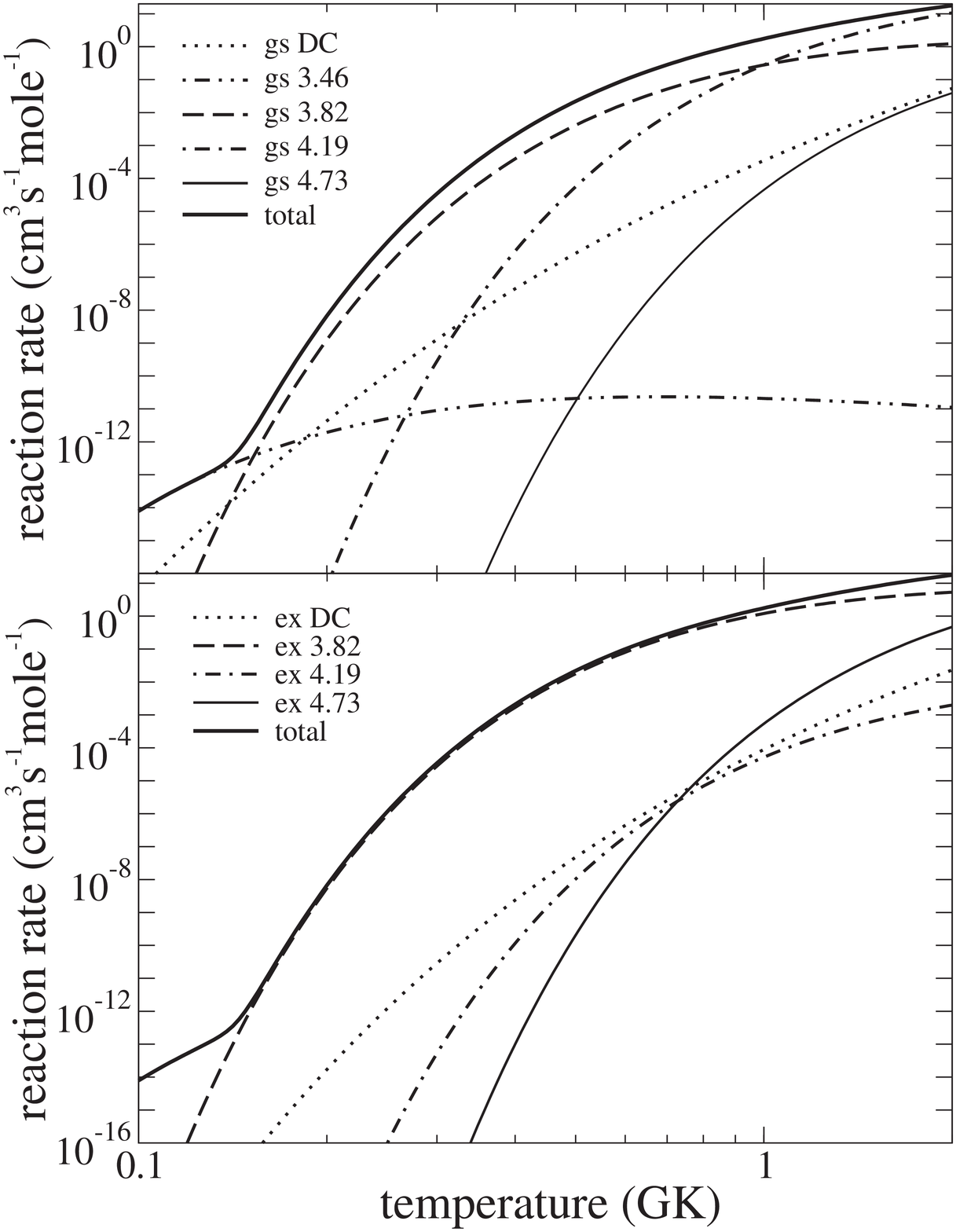}
\caption{\label{FigContr} The contributions of various individual resonances
and through direct capture (DC) to the $^{32}$Cl(p,$\gamma$)$^{33}$Ar
reaction rate as functions of temperature. In the legend, resonances are
labeled with their excitation energy in $^{33}$Ar.
The upper panel shows
contributions from ground state capture, the lower panel contributions from 
capture on the first excited state in $^{32}$Cl in each case weighted with the relative
population of the respective target state. In addition, both panels show the same
total $^{32}$Cl(p,$\gamma$)$^{33}$Ar reaction rate for comparison. 
Contributions from ground state capture via the 3.36 MeV resonance and from 
capture on the first excited state via the 3.46 MeV resonance are too 
small to appear in this graph.}

\end{figure}

The contributions from individual resonances and direct capture to the
total $^{32}$Cl(p,$\gamma$)$^{33}$Ar reaction rate are shown in Fig.~\ref{FigContr}.
The temperature dependent relative population of the ground and first excited
state in $^{32}$Cl has been taken into account.
Direct capture on
the ground or the excited state
is negligible over the entire relevant temperature range of 0.2-2~GK.

As Fig.~\ref{FigContr} shows, the capture rate contributions from 
ground and excited target states vary greatly as $\Gamma_p$ depends
strongly on proton energy, spin, and nuclear structure. 
For $^{32}$Cl(p,$\gamma$)$^{33}$Ar the most dramatic change
occurs for the 
$5/2^+$ 3.819 MeV resonance. For this resonance 
the spin of the excited state ($2^+$) is one unit larger than for the 
ground state (1$^+$) allowing for s-wave protons to populate the 
$5/2^+$ resonance in addition to d-wave protons. 
As a result the proton width for capture on the excited state is a factor of 4.3 
larger than for ground state capture despite the slightly lower proton energy. 
As a consequence (see Eq. \ref{EqTotal}) the
resonant capture on the first excited state in $^{32}$Cl via the 3.819 MeV
$5/2^+$ state in $^{33}$Ar becomes the dominant contribution
to the stellar $^{32}$Cl(p,$\gamma$)$^{33}$Ar reaction rate for all relevant temperatures
between 0.15 and 2 GK (see Fig.~\ref{FigContr}). This is despite the 
fact that the population of the first excited state in $^{32}$Cl for 
example at 0.3~GK is only 5\%. As Eq.~\ref{EqTotal} demonstrates, the
relative importance of contributions from capture on 
various states in the target is largely independent of 
their population (with the exception of their contribution to the 
partition function), what matters most is the proton width for the transition 
into the resonant state. 

The remaining uncertainty from resonance energies ranges from 20 - 50\% for the 
most important temperatures between 0.3 and 0.8~GK and increases to about a factor of 2
at 1.5~GK. It 
originates mainly from the small experimental uncertainties in the $^{33}$Ar excitation 
energies and the reaction Q-value. For the experimentally unknown 
higher lying levels we assume an uncertainty in the excitation energy of 100~keV,
which is a typical deviation for shell model predictions of Coulomb shifts in this 
region. These levels only begin to contribute to the total reaction rate beyond around 1~GK
and lead to the somewhat increased error at highest temperatures. 
At this level of error, other uncertainties in the shell model calculations now become
relevant as well. 
The Woods-Saxon well used for the proton scattering widths
is adjusted so that the total density for protons filled up to the
Fermi energy reproduces the rms charge radius
of nearby stable nuclei. We estimate an uncertainty of 5\% 
in the Woods-Saxon radius paramter which leads to a 15\% uncertainty
in the proton decay widths. 

In principle one could 
estimate the uncertainty of the shell model spectroscopic factors  
by comparing calculated spectroscopic factors with experimental 
data. No experimental data are available for the here relevant proton spectroscopic 
factors in $^{33}$Ar or the corresponding neutron spectroscopic 
factors in the $^{33}$P mirror. In addition,
small spectroscopic
factors are in general difficult to reliably extract from direct reaction cross
sections due to the possibility of multi-step routes.
However, we can test our
shell model descriptions for the relevant states in $^{33}$Ar by comparing their 
relatively large
neutron spectroscopic factors for $^{34}$Ar to $^{33}$Ar neutron removal 
with experimental data for the mirror proton spectroscopic factors
for $^{34}$S to $^{33}$P proton removal. We limit the discussion to the 
5/2$^+$ states, as the $^{32}$Ar(p,$\gamma$)$^{33}$Ar is dominated by the 
resonant contribution from 
the third 5/2$^+$ state in $^{33}$Ar.
The calculations and experimental data for the 
5/2$^+$ states are summarized in Tab.~\ref{TabSpec_n}. 
The calculations are carried out with the original USD interaction as well 
as close to final version of a new sd-shell interaction USD05 \cite{usd05}
obtained from a least squares fit to about 600 levels in the sd-shell nuclei.
The spectroscopic factors are all large numbers and in good agreement with the experimental
data from the mirror reaction. When calculated levels can be
matched to experimental levels with the present degree of accuracy (about
100 keV) the comparison in Tab.~\ref{TabSpec_n} shows that the largest calculated
spectroscopic factors are accurate to about 20\% and the moderately
large spectroscopic factors (as for the second 5/2$^+$ state) are accurate
to about a factor of two. 

\begin{table}
\caption{\label{TabSpec_n} Spectroscopic factors $C^2S$ for d$_{5/2}$ neutron removal from $^{34}$Ar 
to various 5/2$^+$ states in $^{33}$Ar calculated with the shell model effective interactions USD and USD05 and 
compared to experimental data \protect\cite{KMK88} from the $^{34}$S to $^{33}$P mirror reaction for d$_{5/2}$ protons.
The excitation energies $E_x$ of the 5/2$^+$ states are the experimentally known 
ones in the $^{33}$P mirror.}
\begin{ruledtabular}
\begin{tabular}{cccc}
$E_x$ (MeV)  & $C^2S$   & $C^2S$   & $C^2S$ \\
Exp          & Exp      & USD05    & USD  \\
1.85         &  1.09    & 0.99     & 1.09 \\
3.49         &  0.31    & 0.27     & 0.70 \\
4.05         &  1.28    & 1.62     & 1.30 \\
5.05         &  1.66    & 1.51     & 1.33 \\
\end{tabular}
\end{ruledtabular}
\end{table}

In addition, we can look at variations in spectroscopic factors calculated 
with different shell model effective interactions.
The spectroscopic factors for the important $^{32}$Cl (2$^+$) to $^{33}$P (5/2$^+$) $l=0$
transitions are shown in Tab.~\ref{TabSpec_p} for the old USD interaction as well as
the new USD05 interaction. All of the spectroscopic
factors are small since the s$_{1/2}$ orbit is mostly filled in $^{32}$Cl.
The $l=0$ spectroscopic factor summed over all final 5/2$^+$ states is
only 0.131 for USD05.
The comparison of the two interactions in Table 2 shows that the
largest of the spectroscopic factors (including the one to the
third 5/2$^+$ state of interest) agree to about 20 \%. 
Spectroscopic factors calculated with other effective 
interactions such as SDPOTA \cite{sdpota} or CWH \cite{cwh}, which
also do a reasonable job of reproducing the energy levels for nuclei in the
upper sd-shell, show a similar behavior. In general
a small spectrosopic factor in the presence of other states
with much larger 
spectroscopic factors may be very uncertain (as observed 
for many of the states in Tab.~\ref{TabSpec_p}). However, when the 
spectroscopic factor is large compared to those for
nearby states it
can be relatively accurate. Thus the
estimated theoretical error for small spectroscopic 
factors must be treated on 
a case by case basis. 

\begin{table}
\caption{\label{TabSpec_p} $l=0$ spectroscopic factors for the 5/2$^+$ states in
$^{33}$Ar for proton capture on the 2$^+$ state in 
$^{32}$Cl. Spectroscopic factors $C^2S$ and excitation energies $E_x$ calculated
with the USD and the USD05 interaction are compared.}
\begin{ruledtabular}
\begin{tabular}{cccc}
$E_x$ (MeV)  & $E_x$ (MeV) & $C^2S$   & $C^2S$ \\
USD05        & USD         & USD05    & USD  \\
1.99   & 2.00  &  0.0013  & 0.0006  \\
3.42   & 3.83  &  0.0022  & 0.012   \\
4.22   & 4.19  &  0.025   & 0.023   \\
4.82   & 5.00  &  0.049   & 0.040   \\
6.39   & 6.64  &  0.00002 & 0.0024  \\
6.54   & 6.93  &  0.017   & 0.0024  \\
\end{tabular}
\end{ruledtabular}
\end{table}

\begin{figure}[ht]
\includegraphics[angle=270,width=8cm]{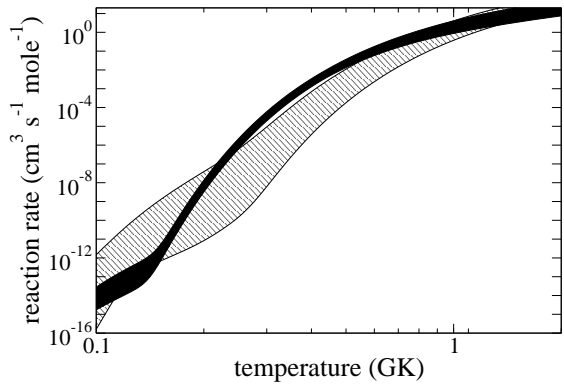}
\caption{\label{FigRate} Total $^{32}$Cl(p,$\gamma$)$^{33}$Ar reaction rate 
from this work (solid area) and previous shell model rate without 
experimental data (hatched area). The width of the area reflects the
estimated uncertainty.}
\end{figure}

To summarize, the shell model can describe the 
neutron single particle strength of the 
here relevant states within 20\% to a factor of 2. In addition, 
the important $l=0$ proton spectroscopic factor for the third
5/2$^+$ state is relatively large compared to other
states and the total $l=0$ proton single particle strength, and its 
calculation should therefore be relatively reliable. We therefore
estimate a factor of 2 uncertainty in the shell model spectroscopic 
factor in this specific case. To take into account other uncertainties
such as the proton single particle widths discussed above, 
we adopt a total uncertainty of a factor of 2.5 for the $^{32}$Cl(p,$\gamma$)$^{33}$Ar
reaction rate on top of the uncertainty from the resonance energies, which 
is accurately calculated as a function of temperature. 
The new reaction rate with its uncertainties is tabulated in Tab.~\ref{TabRat} and 
displayed in Fig.~\ref{FigRate}. The total estimated uncertainty of the rate is
about a factor of 3 at temperatures around 0.3-0.7 GK to about a factor of
6 at temperatures around 1.5~GK. 

\begin{table}
\caption{\label{TabRat}Reaction rate $N_A<\sigma v>$ as a function of temperature $T$. Given 
             is the recommended rate from this work (rec) as well as a lower and upper rate that reflect the estimated
             error bar.}
\begin{ruledtabular}
\begin{tabular}{cccc}
$T$ (GK)  & \multicolumn{3}{c}{$N_A<\sigma v>$ (cm$^3$/s/mole)} \\ 
 & rec & lower & upper \\ 
0.1 &  7.85e-15 &  1.73e-15 &  2.81e-14 \\ 
0.2 &  6.62e-09 &  3.20e-09 &  1.36e-08 \\ 
0.3 &  3.44e-05 &  1.94e-05 &  6.04e-05 \\ 
0.4 &  2.09e-03 &  1.27e-03 &  3.41e-03 \\ 
0.5 &  2.21e-02 &  1.33e-02 &  3.67e-02 \\ 
0.6 &  9.97e-02 &  5.80e-02 &  1.77e-01 \\ 
0.7 &  2.82e-01 &  1.58e-01 &  5.37e-01 \\ 
0.8 &  6.06e-01 &  3.27e-01 &  1.24e+00 \\ 
0.9 &  1.09e+00 &  5.67e-01 &  2.39e+00 \\ 
1.0 &  1.76e+00 &  8.75e-01 &  4.06e+00 \\ 
1.5 &  7.94e+00 &  3.40e+00 &  2.03e+01 \\ 
2.0 &  1.78e+01 &  7.48e+00 &  4.39e+01 \\ 
\end{tabular}
\end{ruledtabular}
\end{table}

Often, reaction rates
are implemented into reaction network codes using a fit of the form:
\begin{equation} \label{EqFit}
N_A <\sigma v> = \sum_i {\rm exp}(a_{0i}+a_{1i} T_9^{-1}+a_{2i} T_9^{-1/3}+a_{3i} T_9^{1/3}
+a_{4i} T_9+a_{5i} T_9^{5/3}+a_{6i} \rm{ln} T_9)
\end{equation}
which is for example used in the {\verb reaclib } reaction rate library. In 
Tab.~\ref{TabFit} we give fit parameters $a_{ni}$ for our new recommended 
$^{32}$Cl(p,$\gamma$)$^{33}$Ar rate as well as for the upper and lower
uncertainties. The fits are accurate to 5\% and 
are valid between 0.1 and 10~GK. Beyond 2 GK we use the Hauser-Feshbach 
predictions from NON-SMOKER \cite{RaT00} using the procedure described 
by the NACRE collaboration \cite{AAR99}. The low temperature behavior is reasonable, so 
the fits can be used in model simulations that encounter lower temperatures. 

\begin{table}
\caption{Fit coefficents (see Eq.~\protect\ref{EqFit}) for the recommended rate from this work (rec)
             as well as for the lower and upper limits that reflect the estimated
             error bar. \label{TabFit}}
\begin{ruledtabular}
\begin{tabular}{crrrrrrr}
     & $a_0$ & $a_1$ & $a_2$ & $a_3$ & $a_4$ & $a_4$ & $a_7$ \\
 rec & 0.150730E+01 & -0.539842E+01 & -0.153745E+02 & 0.187918E+02 &
       0.126139E+01 & -0.208373E+00 & -0.134957E+02 \\
     & 0.542869E+02 & -0.782163E+01 & 0.367513E+03 & -0.425108E+03 &
       0.502875E+01 & 0.823457E+00 & 0.262097E+03 \\ \hline
lower & 0.551123E+03 & -0.158099E+02 & 0.886227E+03 & -0.150891E+04 &
        0.878013E+02 & -0.495734E+01 & 0.714263E+03\\
     &  -0.517996E+02 & -0.156901E+01 & -0.173733E+03 & 0.237842E+03 &
        -0.114259E+02 & 0.538535E+00  & -0.124462E+03 \\ \hline
upper & 0.299493E+03 & -0.143477E+02 & 0.705984E+03 & -0.104556E+04 &
        0.530983E+02 & -0.269508E+01 & 0.533394E+03\\
      & -0.269317E+03 & -0.794584E+00 & -0.279531E+03 & 0.597651E+03 &
        -0.505961E+02 & 0.399204E+01 & -0.244496E+03\\
\end{tabular}
\end{ruledtabular}
\end{table}

\section{Discussion}

\begin{figure}[ht]
\includegraphics[width=6cm,angle=270]{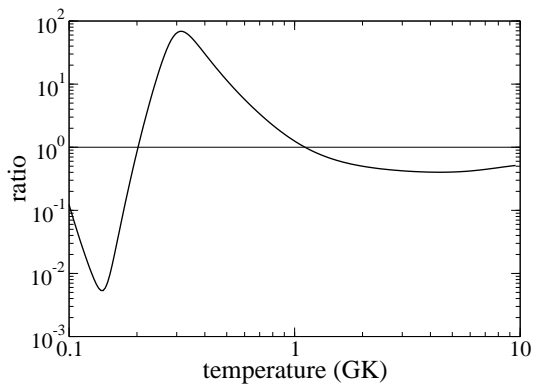}
\caption{\label{FigRatio} Ratio of the $^{32}$Cl(p,$\gamma$)$^{33}$Ar reaction rate of 
this work to the rate calculated by Herndl et al. \protect\cite{HGW95}. }
\end{figure}

Our ground state capture rate for $^{32}$Cl(p,$\gamma$)$^{33}$Ar is similar to the rate 
given by Clement et al. \cite{CBB04}. Small differences arise from the slightly 
improved calculation of the proton single particle widths, the
comprehensive reevaluation
of the shell model uncertainties, and the modifications in the proton widths by 
taking into account transitions into excited target states. The major change however comes
from the consideration of the capture on the first excited state in $^{32}$Cl. 
In the critical temperature range around 0.2-0.4 GK where for typical X-ray
burst conditions (densities of 10$^{5-6}$~g/cm$^3$ and hydrogen mass fractions of 0.1-0.7)
the reaction rate becomes comparable to the burst timescales (10-100 s)
our rate is a factor of more than 4 larger than the ground state capture rate shown
in Clement et al. \cite{CBB04}. In the same temperature range our rate is
up to a factor of 70 larger than the recommended shell model based rate in Herndl et al.
\cite{HGW95} (see Fig. ~\ref{FigRatio}).

\begin{figure}[ht]
\includegraphics[width=6cm,angle=270]{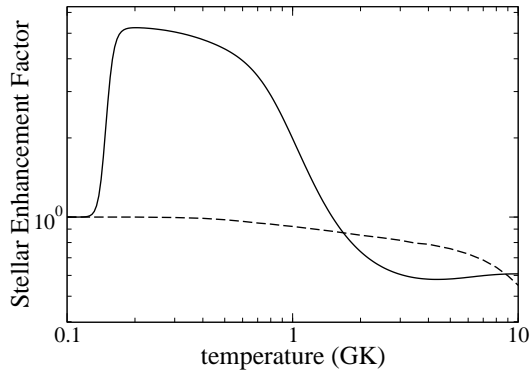}
\caption{\label{FigSEF} Stellar enhancement factor derived in this work (solid line)
and from the statistical model code NON-SMOKER (dashed line).}
\end{figure}

The impact of thermally excited states in the target nucleus on a reaction
rate is often expressed in terms of a "Stellar Enhancement Factor", or SEF, which 
is defined as the ratio of the actual capture rate to the ground state 
capture rate. Fig.~\ref{FigSEF} shows the SEF determined in this work. 
The 
enhanced proton capture rate from the first excited state in $^{32}$Cl leads
to a dramatic enhancement of the reaction rate of up to a factor of 5 at the 
important temperatures around 0.2-0.4~GK. 
Recent reaction rate compilations
typically use SEFs that are calculated with the statistical Hauser-Feshbach 
model \cite{AAR99,Iliadis01}. For comparison we also show in Fig.~\ref{FigSEF}
the SEF obtained with the Hauser-Feshbach code NON-SMOKER \cite{RaT00}. Clearly 
in this case a statistical approach to calculating the SEF does not give 
the correct result, even in the temperature range above 0.7~GK where strictly 
speaking the Hauser-Feshbach model is expected to be applicable \cite{RTK97,RaT00}. This 
is no surprise as
here the SEF is entirely dominated by the properties of a single resonance. However, 
this example illustrates that large uncertainties can be introduced
when using statistical model SEFs far from stability when reaction rates are
dominated by a few resonances only. A better general approach would be to calculate SEFs
within the shell model.

Clearly accurate masses and excitation 
energies are the single most important step for improving the accuracy of theoretical 
reaction rates. 
For comparison Fig.~\ref{FigRate} also shows the uncertainty 
band of the reaction rate for the case that none of the excitation energies 
would be known experimentally, a situation that is very common along 
the rp-process path.  Clearly, without 
experimental excitation energies reaction rates in the rp-process far from 
stability can be uncertain by many orders of magnitude. A measurement 
of excitation energies and Q-values to better than 10~keV can reduce this 
uncertainty to about a factor of 3.

To obtain in critical cases even more precise reaction rates one would need to perform 
indirect measurements of spectroscopic factors, or perform a direct measurement
of the reaction rate 
at astrophysical energies. Note, however, that in this particular case
a direct measurement would not be possible as the reaction rate is 
dominated by capture on the first excited state in the $^{32}$Cl target 
nucleus. This underlines the importance of indirect methods in determining 
accurate stellar reaction rates.

We thank M. Wiescher for pointing out the importance of the timescale for thermal excitation
and O. Sorlin for pointing out the potential relevance of the first excited state in 
$^{32}$Cl.
This work was supported by NSF grants PHY 0110253 (NSCL) and PHY 0216783 (Joint Institute for 
Nuclear Astrophysics). H.S. acknowledges support through the Alfred P. Sloan 
Foundation. B.A.B. acknowledges support from 
NSF grant PHY-0244453. 
C.A.B. thanks for support by the U.S. Department of Energy under grant
No. DE-FG02-04ER41338.

\end{document}